%% file: main.tex
\documentclass[lettersize,journal]{IEEEtran}
\def\BibTeX{{\rm B\kern-.05em{\sc i\kern-.025em b}\kern-.08em
    T\kern-.1667em\lower.7ex\hbox{E}\kern-.125emX}}

\IEEEoverridecommandlockouts

\def\BibTeX{{\rm B\kern-.05em{\sc i\kern-.025em b}\kern-.08em
    T\kern-.1667em\lower.7ex\hbox{E}\kern-.125emX}}
\usepackage{mathtools}
\usepackage[nodisplayskipstretch]{setspace}
\usepackage[utf8x]{inputenc}

\DeclareMathOperator*{\argmax}{arg\,max}

\newcommand\bl[1]{\bold{#1}}
\newcommand\wc{\bold{w}_\textsf{c}}
\newcommand\wct{\bold{w}_\textsf{c}(t)}
\newcommand\rw{\bold{w}_\textsf{r}}
\newcommand\wrt{\bold{w}_\textsf{r}(t)}

\usepackage{subfigure}
\usepackage{moreverb}
\usepackage{epsfig}
\usepackage{amsmath,amssymb,bm,amsthm,mathrsfs,dsfont}
\usepackage{fancyhdr}
\usepackage{adjustbox,lipsum}
\usepackage[font={small}]{caption}
\usepackage{color,soul}
\usepackage{graphicx}
\usepackage{amsfonts}
\usepackage{cite}
\usepackage{microtype}
\usepackage[colorlinks,bookmarksopen,bookmarksnumbered,citecolor=blue,urlcolor=red]{hyperref}

\usepackage{nomencl}
\makenomenclature

\usepackage{algorithmicx}
\usepackage{algpseudocode}
\usepackage[norelsize, linesnumbered, ruled, lined, boxed, commentsnumbered]{algorithm2e}

\allowdisplaybreaks

\begin{document}
\title{ 
Dynamic Joint Communications and Sensing Precoding Design: A Lyapunov Approach

}
\author{
 Abolfazl~Zakeri,~\textit{Member, IEEE}, Nhan~Thanh~Nguyen,~\textit{Member, IEEE},
 Ahmed Alkhateeb,~\textit{Senior Member, IEEE},
 and Markku~Juntti,~\textit{Fellow, IEEE}
 \vspace{-2 em }
 \thanks{A. Zakeri, N. T. Nguyen, and M. Juntti are with the Centre for Wireless
Communications (CWC), University of Oulu, Oulu 90014, Finland, Emails:
\{abolfazl.zakeri; nhan.nguyen; markku.juntti\}@oulu.fi. Ahmed Alkhateeb is with the School of Electrical, Computer, and Energy Engineering, Arizona State University, Email: alkhateeb@asu.edu
 }
}
\maketitle
	\begin{abstract}
This letter proposes a dynamic joint communications and sensing (JCAS) framework to adaptively design dedicated sensing and communications precoders. We first formulate a stochastic control problem to maximize the long-term average signal-to-noise ratio for sensing, subject to a minimum average communications signal-to-interference-plus-noise ratio requirement and a power budget. Using Lyapunov optimization, specifically the drift-plus-penalty method, we cast the problem into a sequence of per-slot non-convex problems. To solve these problems, we develop a successive convex approximation method. Additionally, we derive a closed-form solution to the per-slot problems based on the notion of zero-forcing.
Numerical evaluations demonstrate the efficacy of the proposed methods and highlight their superiority compared to a baseline method based on conventional design.
\end{abstract}
\input{Chapters/Introduction}
\input{Chapters/SystemModel_ProbFor}

 \input{Chapters/Solution}

\input{Chapters/Simulations}
\input{Chapters/Conclusion}

\bibliographystyle{ieeetr}
\bibliography{Bib_References/conf_short,
Bib_References/IEEEabrv,
Bib_References/Bibliography}

\end{document}

%% file: Chapters/Introduction.tex
\section{Introduction} 
The coexistence of communications and sensing has been increasingly attracting attention from academia and industry due to its profound potential for resource efficiency, low cost, and hardware convergence \cite{Eldar_mSP}.
Existing studies in joint communications and sensing (JCAS) can be broadly categorized into radar-centric, communication-centric, and joint design approaches \cite{Nhan_2024PerformanceAA}.
Several interesting relevant problems were addressed, such as waveform design \cite{jsac_wfd_MH} and transmit beamforming design \cite{waveformdesing_IT}, under various performance metrics. Notably, these include the signal-to-clutter-plus-noise ratio \cite{scnr_jsac} and the Cramér–Rao lower bound \cite{Liu_cram2022} for sensing, and quality of service (QoS) constraints, such as a minimum signal-to-interference-plus-noise ratio (SINR) \cite{Ata_2024AdvancedID}, among others for the communications.
\\\indent 
Most existing studies in JCAS however focus on optimizing the system for a given time, e.g., \cite{WenCai2023,jsac_wfd_MH,scnr_jsac,Ata_2024AdvancedID}, regardless of potential variations in network load, resource availability, and memory-dependent time-varying parameters such as wireless channel characteristics \cite{ribeiro2025mobility} and impulse responses of sensed targets~\cite{10600135}. At the same time, satisfying instantaneous (per-slot) constraints for radar and communications might be challenging, if not impossible. Yet, the system can still operate if average radar and communications requirements are met; this relates to the non-guaranteed bit rate within the QoS requirements in 3GPP Release 19 \cite{3gpp_23-501}.
Besides, there are inherent differences between radar and communications operations, depending on the time variations. For instance, the demand for communications is generally lower at night compared to daytime, whereas many sensing applications require continuous operation throughout the day and night. 
\\\indent
Motivated by the above facts, and to account for a dynamic balance in the resource utilization between radar and communications, this study addresses a JCAS problem aiming to \textit{dynamically} design \textit{dedicated} precoders for radar and communications to satisfy an average SINR requirement for communications and maximize average radar SNR.
\\\indent
We consider a JCAS system with a base station (BS), a radar target, and a single-antenna user, where communications occur in the downlink. At each time slot, the BS simultaneously transmits data to the user and receives the echo signal back from the target. The goal is to dynamically design dedicated precoders for both radar and communications.
We formulate a stochastic control problem to maximize the long-term average radar SNR, subject to an average SINR constraint for communications and the BS’s power budget.
The main challenge here is to design a dynamic precoding algorithm under uncertainty and time variations while ensuring strict per-slot and average constraints.
Using Lyapunov optimization, we cast the problem into a sequence of per-slot non-convex optimization problems. To solve these per-lot problems, we apply the successive convex approximation (SCA) method. Additionally, we provide a closed-form solution based on zero-forcing (ZF), resulting in a more efficient dynamic algorithm.
Simulation results demonstrate the effectiveness of the proposed algorithms and their performance compared to a baseline algorithm. The results show that the ZF-based method achieves nearly the same performance as the SCA-based method while offering a considerably lower computational complexity. 
\\\indent
A handful of papers studied the problem of dynamic JCAS design, e.g., \cite{Nikbakht2024AMR,jsac_lya_drlglbW,average_jsac_coml,onlinLearnig_jsac,10600135,gursoy_jsac_learning}.
Work \cite{10600135} developed a dynamic power allocation strategy for a vehicular setup to maximize the average mutual information for the radar, subject to minimum average rate and power budget constraints.
The work in \cite{average_jsac_coml} focused on maximizing average throughput under a hybrid repeat-request protocol, with constraints on minimum average probability of detection and average power utilization.
Reference \cite{onlinLearnig_jsac} proposed a learning-based algorithm for dynamic resource management, while \cite{Nikbakht2024AMR} introduced a memory-based learning method for online channel sensing using backscattered signals.
The study in \cite{jsac_lya_drlglbW} combined Lyapunov optimization with deep reinforcement learning to provide dynamic power and subcarrier allocation, subject to average data rate and sensing requirements.
The work by \cite{gursoy_jsac_learning} proposed a deep reinforcement learning method to dynamically allocate dwell time for target tracking and data transmission. 
\\\indent
Unlike previous works, we devise a dynamic, dedicated radar and communications precoding algorithm aimed at maximizing the average sensing SNR, subject to constraints on average communications SINR and total power budget of the BS. The most closely related work to this paper is \cite{10600135}, which also employs Lyapunov optimization to address a time-averaged problem. However, in contrast to \cite{10600135}, we introduce the SCA method and a ZF-based approach with a closed-form solution to the per-slot problems. Additionally, our problem with the precoding design fundamentally differs from the dynamic power allocation problem studied in \cite{10600135}.

%% file: Chapters/SystemModel_ProbFor.tex
\section{System Model and Problem Formulation}
\allowdisplaybreaks
We consider a JCAS system with a BS, a target, and a single-antenna (communications) user. The communications takes place on the downlink. The BS is equipped with $N$ transmit antennas and $M$ receive antennas. For clarity of the presentation, we use subscript ``$\textsf{r}$" for notations associated with radar/sensing, and ``$\textsf{c}$" for those with communications. 
%
At time slot $t$, the BS simultaneously transmits probing signals to the target and data signals to the communications user using dedicated precoding vectors $\wrt \in\Bbb{C}^{N\times 1}$ and ${\wct \in\Bbb{C}^{N\times 1}}$, respectively. The BS then receives the radar echo signal using $M$ antennas.
\\\indent
Let $\bl{h}(t)\in\Bbb{C}^{N\times 1}$ denote the channel from the BS to the user in slot $t$. Note that, due to fluctuations in the wireless channels, $\bl{h}(t)$ is essentially a stochastic process.
The received signal by the user is then given by 
\begin{equation}
    y_{\textsf{c}}(t) = \bl{h}^{\textsf{H}}(t)\wct {s}_{\textsf{c}}(t)+\bl{h}^{\textsf{H}}(t)\wrt {s}_{\textsf{r}}(t)+n_{\textsf{c}}(t),
\end{equation}
where ${s}_{\textsf{c}}(t)$ is the transmit (data) signal to the communications user, $\Bbb{E}\{|{{s}_{\textsf{c}}(t)}|^2\}=1$,
${s}_{\textsf{r}}(t)$ is the transmit radar signal with unit energy,
and $n_{\textsf{c}}(t)\in \Bbb{C}$ is additive white Gaussian noise (AWGN) following the distribution $\mathcal{CN}(0,\sigma^2_{\textsf{c}})$, with $\sigma^2_{\textsf{c}}$ denoting the noise variance at the communications user.
The radar's dedicated signal becomes interference for the user. As such,  the SINR at the user at slot $t$ is given by
\begin{equation} 
\gamma_{\textsf{c}}(t) = \frac{|\bl{h}^{\textsf{H}}(t) \wct|^2}{|\bl{h}^{\textsf{H}}(t) \wrt|^2 + \sigma_{\textsf{c}}^2}. 
\end{equation} 

While transmitting communications signals to the users, the
BS also receives echo signals from the target.
Let ${\bl{G}(t)\in\Bbb{C}^{M\times N}}$ denote 
the two-way channel between the BS and the target in slot~$t$. We model $\bl{G}(t)$ as \cite{WenCai2023}
\begin{equation}\label{eq_radar_channel}
\bl{G}(t) = \bold{b}(t)\bold{a}^{\textsf{H}}(t),
\end{equation}
where $\bl{a}(t)\in\Bbb{C}^{ N\times 1 }$ and $\bl{b}(t)\in\Bbb{C}^{ M\times 1 }$ are the steering vectors of the transmit and receive antennas, respectively.
The received radar sensing signal at the BS is given by
\begin{equation}
    \bold{y}_{\textsf{r}}(t)= \alpha \bl{G}(t)\wct {s}_{\textsf{c}}(t) + \alpha \bl{G}(t)\wrt {s}_{\textsf{r}}(t) + \bold{n}_{\textsf{r}}(t),
\end{equation}
where $\alpha$ is the complex-valued reflection coefficient \cite{WenCai2023},\footnote{Notice that estimating the phase of $\alpha$ is challenging. However, as it becomes clear next, our analysis depends solely on its magnitude.} and
${ \bold{n}_{\textsf{r}}(t)\in\Bbb{C}^{M\times 1} }$ is an AWGN noise vector whose entries are drawn independently from complex Gaussian distribution $\mathcal{CN}(0,\sigma_{\textsf{r}}^2)$, with $\sigma_{\textsf{r}}^2$ being the noise power at the BS's receiver.
The communications beamformer $\wct$ is assumed to also benefit the radar, e.g., \cite{Ata_2024AdvancedID}. Accordingly, the SNR of the received echo signal at the BS is given by
\begin{equation} 
\gamma_{\textsf{r}}(t) = \dfrac{ |\alpha|^2 \left(\|\bl{G}(t)\wrt\|^2 + \|\bl{G}(t)\wct \|^2 \right)}{ \sigma_{\textsf{r}}^2 }.
\end{equation} 
\textit{Problem Formulation:}~Let us define the time average expected radar SNR and communications SINR  as
\begin{align}
  &  \bar {\gamma}_{\textsf{r}} = \limsup_{T\rightarrow \infty } \dfrac{1}{T} \sum_{t=1}^{T} \mathbb{E} \{\gamma_\textsf{r} (t)\},
    \\
    &
     \bar {\gamma}_{\textsf{c}} = \limsup_{T\rightarrow \infty } \dfrac{1}{T} \sum_{t=1}^{T} \mathbb{E} \{\gamma_\textsf{c} (t)\},
\end{align}
where $\Bbb{E}\{\cdot\}$ is the expectation taken with respect to the system randomness over time. 
At each slot, we aim to find the best precoding vectors $\wct$ and $\wrt$ that optimize the average SNR for the radar while ensuring a minimum average SINR for the user, subject to the BS's power budget. Formally, our goal is to solve the following stochastic control problem:
\begin{subequations}
       \label{op_1}
       \begin{align}
     \underset{\{\bl{w}_{\textsf{c}}(t),\bl{w}_{\textsf{r}}(t)\}_{t=1,2,\ldots}}{\mbox{maximize}}~~~   &
           \bar{\gamma}_{\textsf{r}}
           \\
        		\mbox{subject to}~~~~~~~~ &  
                \label{eq_cons_rate}
                \bar{\gamma}_{\textsf{c}} \geq \gamma_{\mathrm{min}}
                \\ &
                \label{eq_cons_power}
                 \|\bl{w}_{\textsf{r}} (t)\|^2 +  \|\bl{w}_{\textsf{c}}(t)\|^2 \leq P_{\mathrm{max}}, ~\forall\, t
                \end{align}
        		\end{subequations} 
where $\gamma_{\min}$ is the minimum average SINR required for the user and $P_{\max}$ is the (transmit) power budget of the BS.
%
Problem~\eqref{op_1} aims to maximize the average radar performance, while constraint~\eqref{eq_cons_rate} ensures a minimum average data rate for a communications user over time. Notably, unlike most related works that impose strict per-slot resource and QoS constraints, our approach allows for dynamic allocation of resources (i.e., space and power) between radar and communications based on long-term performance requirements. This enables a dynamic power allocation among the two subsystems over the time dimension. Furthermore, it also brings additional flexibility which is more relevant to radar applications and some communications applications, e.g., applications with non-guaranteed bit-rate  QoS flow requirements such as web browsing and video streaming \cite{3gpp_23-501}. 


%% file: Chapters/Solution.tex
\section{Proposed Solution to Problem \eqref{op_1}}
\subsection{Lyapunov Method}
To solve the main stochastic problem in \eqref{op_1}, we derive a dynamic precoding algorithm using the drift-plus-penalty method \cite{Neely_Sch}. 
This algorithm enables the use of convex optimization tools and does not require prior knowledge of the system dynamics or the probabilities associated with the user and radar channels.
The main idea is to enforce the average constraint~\eqref{eq_cons_rate} to a queue stability constraint. 

Let $Q(t)$ denote the virtual queue associated with constraint~\eqref{eq_cons_rate} in slot $t$ which evolves as 
\begin{equation}\label{eq_virtualQueue}
    Q(t+1) = \max [Q(t) - \gamma_{\textsf{c}}(t) ,0] + \gamma_{\mathrm{min}}. 
\end{equation}
The process $Q(t)$ can be seen as a queue with service rate $\gamma_{\textsf{c}}(t)$ and arrival rate $\gamma_{\mathrm{min}}$.
By \cite[Ch. 2]{Neely_Sch}, the time average constraint~\eqref{eq_cons_rate} is satisfied when the queue is strongly stable, i.e., $\limsup_{T\rightarrow \infty} \frac{1}{T}\sum_t^T \Bbb{E}\{Q(t)\} < \infty$. Next, we define
the Lyapunov function and its drift to account for the queue stability and proceed with the drift-plus-penalty method. 

Let $L(Q(t))=\frac{1}{2}Q^2(t)$ be the quadratic Lyapunov function \cite[Ch. 3]{Neely_Sch}. By minimizing the expected change of the Lyapunov function
from one slot to the next, the virtual queue can be stabilized
\cite[Ch. 3]{Neely_Sch}. Let $S(t)\triangleq \{Q(t),\bold{h}(t),\bold{G}(t)\}$ denote the network state
in slot~$t$. The one-slot conditional Lyapunov drift, denoted
by $\Delta(t)$, is the expected change in the Lyapunov
function over one slot given the current system state $S(t)$.
Accordingly, $\Delta(t)$ is defined as \cite[Eq. 3.13]{Neely_Sch}
\begin{equation}\label{eq_drift}
    \Delta(t) = \Bbb{E} \{L(Q(t+1)) - L(Q(t))\,|\,S(t)\}.
\end{equation}
	Applying the drift-plus-penalty method, we need to design $\wct$ and $\wrt$ every slot $t$ that  minimizes a bound on the drift-plus-penalty function 
    \begin{equation}\label{eq_dpp_func}
    \Delta(t) -    V \Bbb{E}\{\gamma_{\textsf{r}}(t)\,|\,S(t)\}  ,
    \end{equation}
    subject to the power constraint \eqref{eq_cons_power}, where $V$ is a non-negative parameter
    chosen to desirably adjust a trade-off between the size of the virtual queue and the objective function of \eqref{op_1}.

Optimizing directly \eqref{eq_dpp_func}  is difficult owing to function $\max[\cdot]$ in the virtual queue evolution in \eqref{eq_virtualQueue}.
Leveraging the fact that for any $Q \ge 0, b \ge 0, A \ge 0$, we have \cite[p. 33]{Neely_Sch}
$$(\max[Q-b,0]+A)^2 \le Q^2+A^2+b^2+2Q(A-b),$$ we can derive the following upper-bound for $\Delta(t)$:
\setlength{\abovedisplayskip}{5pt}
\setlength{\belowdisplayskip}{5pt}
\begin{equation}
    \Delta(t) \le C + Q(t) - \Bbb{E}\{Q(t)\gamma_{\textsf{c}}(t)\,|\,S(t)\},
\end{equation}
where $C$ is a positive constant. 

Following the standard procedure of the drift-plus-penalty method, we use the approach of opportunistically minimizing an expectation to optimize the upper-bound of the drift-plus-penalty function \cite[Ch. 3]{Neely_Sch}. Noting that the constant terms in the drift-plus-penalty do not impact the solution, to obtain our dynamic precoding algorithm, we now aim to solve the following \textit{per-slot} optimization problem for a given~$ S(t) $:
\begin{subequations}
       \label{op_perslot}
       \begin{align}
        \underset{\wc,\rw}{\mbox{maximize}}~~~   &\label{eq_obj_prslot}
        \dfrac{V|\alpha|^2}{\sigma_{\textsf{r}}^2} \left( \|\bl{G}\bl{w}_{\textsf{r}}\|^2 + \|\bl{G}\bl{w}_{\textsf{c}}\|^2 \right)
        +
          Q
         \dfrac{|\bl{h}^{\textsf{H} } \bl{w}_\textsf{c}|^2}{|\bl{h}^{\textsf{H}} \bl{w}_\textsf{r}|^2 + \sigma_{\textsf{c}}^2}
           \\
        		\mbox{subject to}~~~ & \label{eq_cons_power_perslot}
                              \|\bl{w}_{\textsf{r}}\|^2 +  \|\bl{w}_{\textsf{c}}\|^2 \leq P_{\mathrm{max}}, 
                \end{align}
        		\end{subequations}
                 where $Q$ is the virtual queue for the given slot.
Notice that because the above problem is for a given slot~$t$, we dropped the time index $(t)$ from the notations for ease of exposition. In the next section, we present our solution algorithm to \eqref{op_perslot}.
\\\indent
Once the per-slot problem in \eqref{op_perslot} is solved, our dynamic precoding algorithm works as described in Alg.~\ref{alg_dynBEM}. At each slot, the BS updates the channel state information and the virtual queue. It then solves the optimization problem in~\eqref{op_perslot} and the system goes to the next slot.
\begin{algorithm} [t]
   \caption{Dynamic Precoding Algorithm to Problem~\eqref{op_1}}
    \label{alg_dynBEM}
    \SetKwInOut{Inputi}{Initialize}
    \SetKwInOut{run}{RUN}
    \setlength{\AlCapSkip}{1em}
     \SetKwInOut{output}{Output}
     \SetKwComment{Comment}{/*}{ }
     \SetKwRepeat{Do}{do}{while}
    \Inputi{  
    set $t=0$, $V$, and initialize $Q(0)=0$.
    }
    \For{each time slot $t$ }{
    Solve problem~\eqref{op_perslot} using SCA or ZF methods in Section~\ref{sec_dpp_perslot_solution}.
        \\ Update $Q(t+1)$ by \eqref{eq_virtualQueue} and the channels~$\bl{h}$~and~$\bl{G}$. 
    } 
\end{algorithm}
\setlength{\textfloatsep}{0.1cm} 
\\\indent 
Finally, we shall demonstrate that Alg.~\ref{alg_dynBEM} is guaranteed to satisfy the average constraint~\eqref{eq_cons_rate}. Since, for any finite $V$, both \(\mathbb{E}\{L(Q(0))\}\) and the SNR at each time slot remain finite 
 under Alg.~\ref{alg_dynBEM}, it follows that the virtual queue is strongly stable. This, in turn, implies that Alg.~\ref{alg_dynBEM} satisfies constraint~\eqref{eq_cons_rate}.
 \vspace{-2 mm}
\subsection{Solving Problem~\eqref{op_perslot}}\label{sec_dpp_perslot_solution}
Problem~\eqref{op_perslot} is a non-convex optimization problem due to the non-concave objective function, thus it is challenging to solve in general. In what follows, we propose two different efficient algorithms to solve the problem. 
The first method is based on SCA, and the second one is based on the idea of ZF.\footnote{It is worth mentioning that problem~\eqref{op_perslot} can also be solved by the semi-definite relaxation (SDR) method. However, because SDR increases the dimension of the problem, it is computationally expensive.} Notably, the ZF-based method yields a closed-form solution, hence, it is more feasible for practical applications.  
\subsubsection{SCA-Based Solution}~The main idea of this method is to replace the convex functions in the objective function~\eqref{eq_obj_prslot} with surrogate functions and then iteratively solve the resulting convex problem. To effectively address the SINR term in~\eqref{eq_obj_prslot}, we introduce an auxiliary variable~$\beta$ and rewrite the problem~as 
\begin{subequations}
       \label{op_perslot_sca}
       \begin{align}
    \underset{\wc,\rw,\beta}{\mbox{maximize}}      ~~~   &\label{eq_obj_sca}
          \|\bl{G}\bl{w}_{\textsf{r}}\|^2 + \|\bl{G}\bl{w}_{\textsf{c}}\|^2 +
          \Tilde{V} \beta
           \\
        		\mbox{subject to}~~~ &  
                              \|\bl{w}_{\textsf{r}}\|^2 +  \|\bl{w}_{\textsf{c}}\|^2 \leq P_{\mathrm{max}}, 
                              \\ &\label{eq_cons_sinr}
                          \beta \le   \frac{|\bl{h}^{\textsf{H} } \bl{w}_{\textsf{c}}|^2}{|\bl{h}^{\textsf{H}} \bl{w}_{\textsf{r}}|^2 + \sigma_{\textsf{c}}^2},
                          \\&
                          \beta \ge 0,  
                \end{align}
        		\end{subequations}     
                where $\Tilde{V}\triangleq \dfrac{Q\sigma_{\textsf{r}}^2}{V|\alpha|^2}$ is a positive constant.
                Applying SCA, we use the first Taylor expansion of the first two terms in the objective~\eqref{eq_obj_sca} as well as constraint~\eqref{eq_cons_sinr} to convexity the optimization problem. Accordingly, at  iteration $i$, the norms in the objective function~\eqref{eq_obj_sca} are linearly approximated by 
                \begin{equation}\label{eq_Gwr_approx}
                    \|\bl{G}\bl{w}^{(i-1)}_{\textsf{r}}\|^2 
                    + 2\Re \left\{ {\bl{w}^{(i-1)}_{\textsf{r}}}^{\textsf{H}} \bold{G}^{\textsf{H}}\bold{G} 
                    \left(\bold{w}^{(i)}_\textsf{r} - \bold{w}^{(i-1)}_\textsf{r}\right)
                  \right  \},
                \end{equation}
                  \begin{equation}\label{eq_Gwc_approx}
                    \|\bl{G}\bl{w}^{(i-1)}_{\textsf{c}}\|^2 
                    + 2\Re \left\{ {\bl{w}^{(i-1)}_{\textsf{c}}}^{\textsf{H}} \bold{G}^{\textsf{H}}\bold{G} 
                    \left(\bold{w}^{(i)}_\textsf{c} - \bold{w}^{(i-1)}_\textsf{c}\right)
                  \right  \}.
                \end{equation}
                Furthermore, to address constraint~\eqref{eq_cons_sinr}, we first rewrite it as 
                \begin{equation}\label{eq_sinrcons_approx}
               |\bold{h}^{\textsf{H}} \bl{w}_{\textsf{r}}|^2 + \sigma_{\textsf{c}}^2
                \le 
                \frac{
              |\bold{h}^{\textsf{H}} \bl{w}_{\textsf{c}}|^2 +
              |\bold{h}^{\textsf{H}} \bl{w}_{\textsf{r}}|^2 + \sigma_{\textsf{c}}^2
                }
                {\beta + 1},
                \end{equation}
                which is non-convex. To transform it into a convex form, at each iteration $i$, its right-hand side is approximated by the first-order Taylor expansion given by 
                         \begin{equation}
    \begin{aligned}
        & \dfrac{
              |\bold{h}^{\textsf{H}} \bl{w}^{(i-1)}_{\textsf{c}}|^2 +
              |\bold{h}^{\textsf{H}} \bl{w}^{(i-1)}_{\textsf{r}}|^2 + \sigma_{\textsf{c}}^2
                }
                {\beta^{(i-1)} + 1}
       \\& \quad  +
          2\Re{ \left\{ \dfrac{{\bl{w}^{(i-1)}_{\textsf{c}}}^{\textsf{H}}\bl{h}\bl{h}^{\textsf{H}} }
               {\beta^{(i-1)} + 1}
               \left(\bl{w}^{(i)}_{\textsf{c}} - \bl{w}^{(i-1)}_{\textsf{c}}\right)
               \right\} }
          \\
        & \quad +
          2\Re{ \left\{ \dfrac{{\bl{w}^{(i-1)}_{\textsf{r}}}^{\textsf{H}}\bl{h}\bl{h}^{\textsf{H}} }
               {\beta^{(i-1)} + 1}
               \left(\bl{w}^{(i)}_{\textsf{r}} - \bl{w}^{(i-1)}_{\textsf{r}}\right)
               \right\} }
          \\
        & -
          \dfrac{
              |\bold{h}^{\textsf{H}} \bl{w}^{(i-1)}_{\textsf{c}}|^2 +
              |\bold{h}^{\textsf{H}} \bl{w}^{(i-1)}_{\textsf{r}}|^2 + \sigma_{\textsf{c}}^2
                }
                {(\beta^{(i-1)} + 1)^2}\left( \beta^{(i)}-\beta^{(i-1)}\right).
    \end{aligned}
\end{equation}
\allowdisplaybreaks
By plugging the approximations \eqref{eq_Gwr_approx}, \eqref{eq_Gwc_approx}, and \eqref{eq_sinrcons_approx} in problem~\eqref{op_perslot_sca}, the resulting problem is convex, which can be solved using standard convex optimization solvers, e.g., CVX. 
\\\indent
Although the SCA-based solution can potentially achieve a locally optimal solution to the per-slot problem~\eqref{op_perslot}, it typically suffers from slow convergence (i.e., higher runtime complexity) and a strong dependency on the initialization of the optimization variables. To address these challenges and offer a computationally simpler solution, we propose an efficient low-complexity solution based on the ZF method.
\subsubsection{ZF-Based Solution} Here, we apply the ZF method to solve problem~\eqref{op_perslot}. ZF imposes that the \textit{direction} of the radar beamformer $\rw$ should be in the null-space of the user channel~$\bold{h}$.  Let us proceed by defining a projection as
\begin{equation}
    \bold{P}_{\bold{h}}~ \triangleq  \bold{I}_{N} - \frac{\bold{h}\bold{h}^{\textsf{H}} }{\|\bold{h}\|^2} \cdot
\end{equation}
 Then, the vector $\bold{w}_{\textsf{r-dir}} = \bold{P}_{\bold{h}}\bold{y} $, i.e., the direction of $\rw$ (with unit norm),  is orthogonal to the channel $\bold{h}$ for any ${ \bold{y}\in \Bbb{C}^{N\times 1} }$.
 It is evident from problem~\eqref{op_perslot} that an optimal $\bold{w}_{\textsf{r-dir}}$ should maximize $\|\bold{G}\bold{w}_{\textsf{r-dir}}\|^2$. This, by plugging $\bold{P}_{\bold{h}}\bold{y} $ in the norm, implies  
$ { \bold{y}^\star \in\argmax ~\{\bl{y}^{\textsf{H}} \bold{P}_{\bold{h}}^{\textsf{H}} \bl{G}^{\textsf{H}} \bl{G} \bold{P}_{\bold{h}} \bl{y} \} } $. Thus, $\bold{y}^{\star}$ should be in direction with the eigenvector corresponding to the largest eigenvalue of matrix $\bold{P}_{\bold{h}}^{\textsf{H}} \bl{G}^{\textsf{H}} \bl{G} \bold{P}_{\bold{h}}$. Given $\bold{y}^{\star}$,  the optimal direction of the radar beamformer $\rw$ can be obtained while its (optimal) power will be determined next.
\\\indent
Having the direction of $\rw$ obtained, the optimization problem with respect to $\bold{w}_{\textsf{c}}$ can be written as 
\begin{equation}\label{eq_opt_wcstart}
{\bold{w}}^{*}_{\textsf{c}}\in \argmax_{\|\bl{w}_\textsf{c}\|^2 \le P_{\max}-P_{\textsf{r}}}~\|\bl{G}\bl{w}_{\textsf{c}}\|^2 +
          K
         |\bl{h}^{\textsf{H} } \bl{w}_\textsf{c}|^2,
         \end{equation}
         where $K\triangleq \dfrac{Q\sigma^2_{\textsf{r}}}{V\sigma^2_{\textsf{c}}}$ is a positive constant, and $P_{\textsf{r}}$ is the allocated transmit power to radar. 
By factorization of the objective function in~\eqref{eq_opt_wcstart}, an optimal direction of $\wc$, $\bold{w}_{\textsf{c-dir}}$, should maximize $\bold{w}^{\textsf{H}}_{\textsf{c}}\bold{B}\bold{w}_{\textsf{c}}$, where 
${ \bold{B} \triangleq \bold{G}^{\textsf{H}}\bold{G} + K\bold{h}^{\textsf{H}}\bold{h} }$ is a symmetric Hermitian matrix.  This suggests that an optimal direction of the user's beamformer, $\bl{w}_\textsf{c-dir}$, should be aligned with the eigenvector corresponding to a maximum eigenvalue of matrix~$\bold{B}$.
Let $P_{\textsf{c}}$ be the communications power. 
Having the optimal direction of the communications and the radar beamformer determined above,  we now shall find their corresponding optimal powers, i.e., $P^{\star}_{\textsf{r}}$ and $P^{\star}_{\textsf{c}}$, subject to the power budget. 
\\\indent
Under ZF, the objective function~\eqref{eq_obj_prslot} is linear in   $P_{\textsf{r}}$ and $P_{\textsf{c}}$. Moreover, it can be seen that at the optimal point, the power budget constraint~\eqref{eq_cons_power_perslot} must be satisfied with equality, i.e., $ P^{\star}_{\textsf{r}} + P^{\star}_{\textsf{c}} = P_{\max}$. 
Using this fact we can derive the optimal power allocations $P^{\star}_{\textsf{r}}$  according to 
\begin{subequations}
       \label{}
       \begin{align}
          { \underset{P_{\textsf{r}}}{\text{maximize}} }~~~   &\label{}
      P_{\textsf{r}}  \|\bl{G}\bl{w}_{\textsf{r-dir}}\|^2 
      \\ &\nonumber
      +
      (P_{\max}- P_{\textsf{r}})\left\{
      \|\bl{G}\bl{w}_{\textsf{c-dir}}\|^2 
        +
          K
        |\bl{h}^{\textsf{H} } \bl{w}_{\textsf{c-dir}}|^2 \right\}
           \\
        		\mbox{subject to}~~~ & \label{}
                            0 \le  P_{\textsf{r}} \le  P_{\mathrm{max}}, 
                \end{align}
        		\end{subequations}
and  $P_{\textsf{c}}^{\star}$ by the fact that 
$  P^{\star}_{\textsf{c}} = P_{\mathrm{max}} - P^{\star}_{\textsf{r}}$. By ignoring the constant terms in the above problem, i.e., the terms with~$P_{\max}$, $P^{\star}_{\textsf{r}}$ is given by: 
\begin{align*}
 P^{\star}_{\textsf{r}} = &   \begin{cases} 
        P_{\mathrm{max}}, & \! \! \!\text{if}  ~ \|\bl{G}\bl{w}_{\textsf{r-dir}}\|^2 
      \ge  \left\{
      \|\bl{G}\bl{w}_{\textsf{c-dir}}\|^2 
        +
          K
        |\bl{h}^{\textsf{H} } \bl{w}_{\textsf{c-dir}}|^2 \right\}
        \\ 
        0, & \! \! \! {\text{otherwise}},
    \end{cases}
\end{align*}
and finally follows by $P^{\star}_{\textsf{c}} = P_{\mathrm{max}} - P^{\star}_{\textsf{r}}$. Interestingly, this ZF-based solution implies that \textit{only one} of the communications or radar subsystem is active (i.e., it has non-zero power) at a given slot, most likely the communications one. This suggests that, with proper dynamic design of communications systems,
it is possible to achieve satisfactory radar performance or execute specific radar tasks that enhance communications. 
{In other words, the radar sensing automatically benefits from the communications signals reflected back to the BS.}

%% file: Chapters/Simulations.tex
\section{Numerical Results}\label{sec_numres}
This section presents simulation results to demonstrate the effectiveness of the derived policy and the impact of various parameters on performance.  We set ${N=4}$, ${M=2}$,   $\sigma^2_{\textsf{c}}=1$ and $\sigma^2_{\textsf{r}}=1$. Each element of the communications user's channel, $\bl{h}(t)$, is generated according to i.i.d. complex Gaussian distribution with zero mean and variance of $1$ \cite{Liu_cram2022}. For the radar two-way channel in \eqref{eq_radar_channel}, we set $\bold{a}=\dfrac{1}{\sqrt{N}}{\left[1,\cdots, e^{j\pi(N-1)\sin(\theta)} \right]}^{\textsf{T}}$, and $ \bold{b}$ is modeled similarly,
both with half wavelength antenna spacing and $ { \theta=\dfrac{\pi}{8} }$ being the target angle of interest. Moreover, we use a power normalization as ${|\alpha|^2 P_{\max}=1}$~\cite{Eldar_coordinatedBistate}. Additionally, for the SCA algorithm used to solve the per-slot problem, we set the convergence tolerance criterion to \(10^{-3}\). The remaining parameters are provided in the caption of each figure. 

In the following figures, we refer to the proposed SCA- and ZF-based Lyapunov algorithms as \textit{Lya-SCA} and \textit{Lya-ZF}, respectively. As a \textit{baseline}, we consider a method that solves problem~\eqref{op_1} for each slot. Specifically, we use ZF for the radar beamformer direction and a matched filter for the communications beamformer to maximize the communications SINR. A line search is then performed to determine the optimal power allocation that satisfies the SINR constraint~\eqref{eq_cons_rate} in each slot.
Notice that the Lya-ZF and the baseline algorithm have the same computational complexity. In the simulations, we consider a period of $2000$ time slots.

\begin{figure}[t!]
\vspace{-0.5cm}
\hspace{-0.35cm}
\subfigure[Average~communications SINR] 
{
\includegraphics[width=0.255\textwidth]{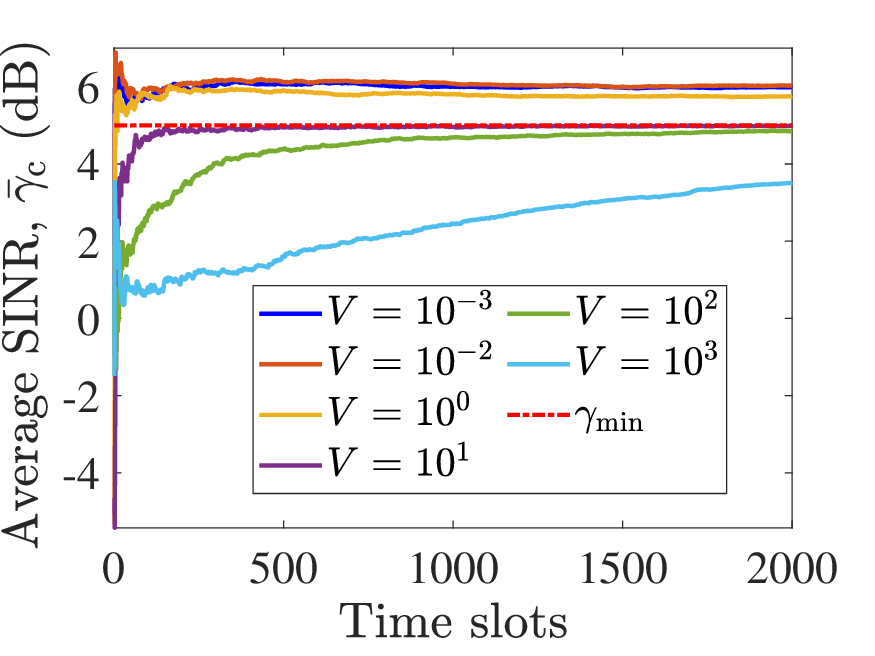}
\label{fig-lyasca_V_C}
}
\hspace{-0.75cm}\subfigure[Average radar SNR]{
\includegraphics[width=0.255\textwidth]{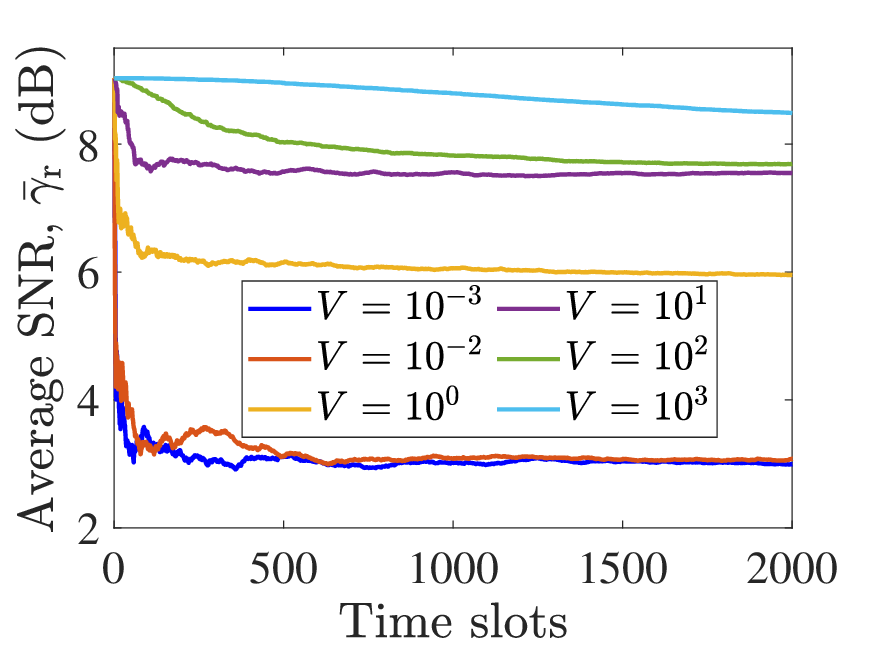 }
\label{fig-lyaSCA_V_R}
}
\caption{Impact of the tradeoff parameter $V$ for the \textit{Lyapunov-SCA} method, for $P_{\max}=0$ dB and $\gamma_{\min}=5$ dB.
}
\label{fig-lyaSCA_V}
\end{figure}

\begin{figure}[t!]
\hspace{-0.4cm}
\subfigure[Average communications SINR] 
{
\includegraphics[width=0.255\textwidth]{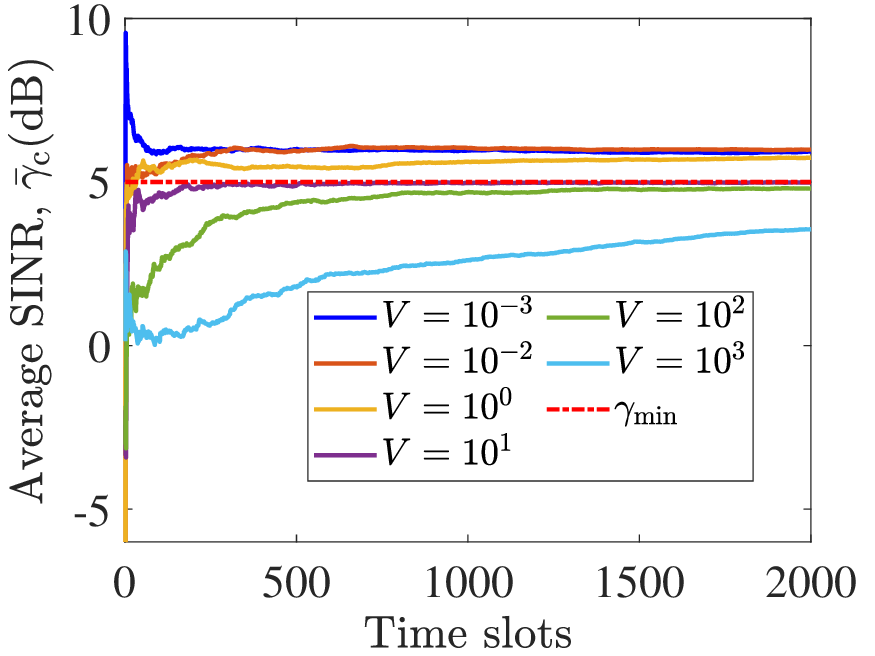}
\label{fig-lyaZf_V_C}
}
\hspace{-0.65cm}\subfigure[Average radar SNR]{
\includegraphics[width=0.255\textwidth]{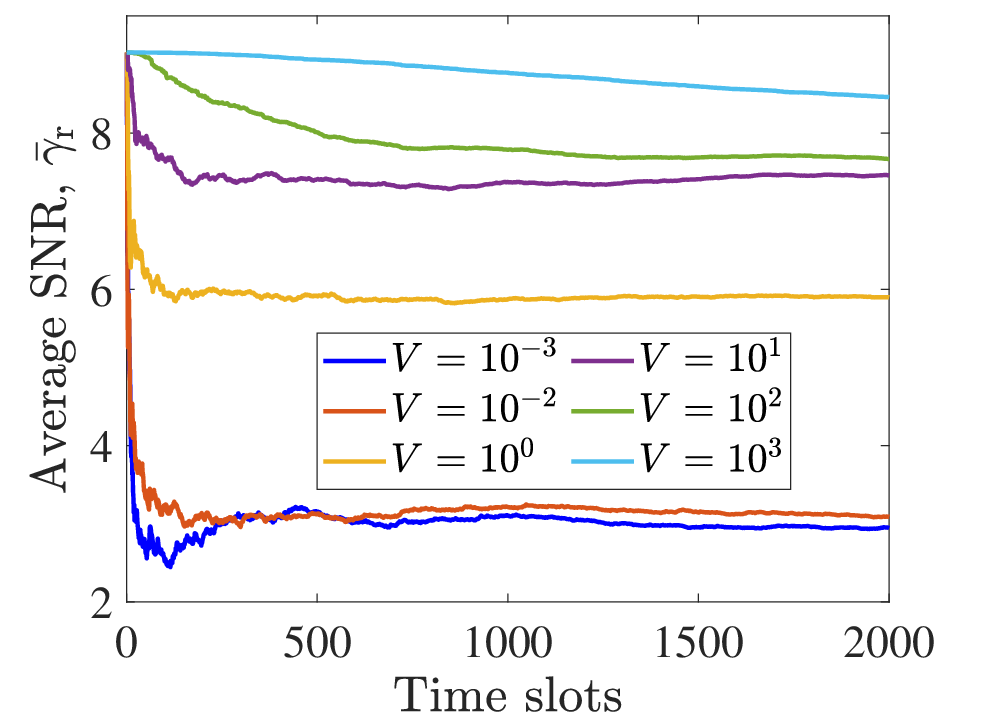 }
\label{fig-lyaZf_V_R}
}
\caption{Impact of the tradeoff parameter $V$ for the \textit{Lyapunov-ZF} method, for $P_{\max}=0$ dB and $\gamma_{\min}=5$ dB.
}
\label{fig-lyaZF_V}
\end{figure}

We illustrate the impact of parameter $V$ and the minimum communications SINR level $\gamma_{\min}$ on the performance of the Lya-SCA algorithm in Fig.~\ref{fig-lyaSCA_V} and the Lya-ZF algorithm in Fig.~\ref{fig-lyaZF_V}. The figures show that both proposed methods satisfy the average constraint~\eqref{eq_cons_rate} for all $V$. However, the convergence speed decreases as $V$ increases. Moreover, as expected, larger values of $V$ yield higher average radar SNR for both algorithms. These observations provide practical insights: A sufficiently large $V$ should be used to enhance radar performance, and increasing $V$ beyond $100$ does not impact the performance.

\begin{figure}[t!]
\hspace{-0.4cm}
\subfigure[${\gamma_{\min}=10}$ dB] 
{
\includegraphics[width=0.255\textwidth]{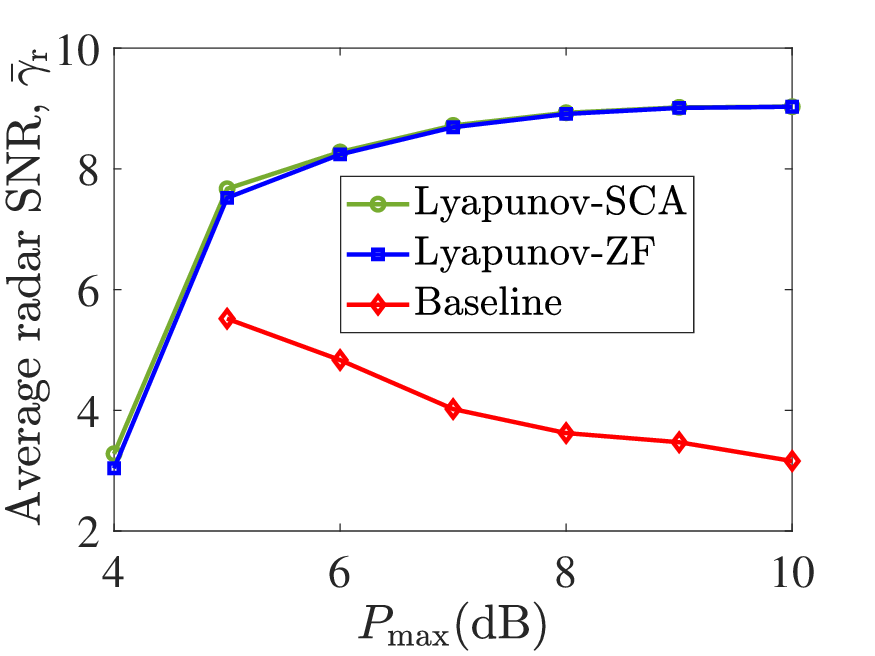}
\label{fig-all_pmax}
}
\subfigure[$P_{\max}=5$~dB]{
\hspace{-0.65cm}\includegraphics[width=0.255\textwidth]{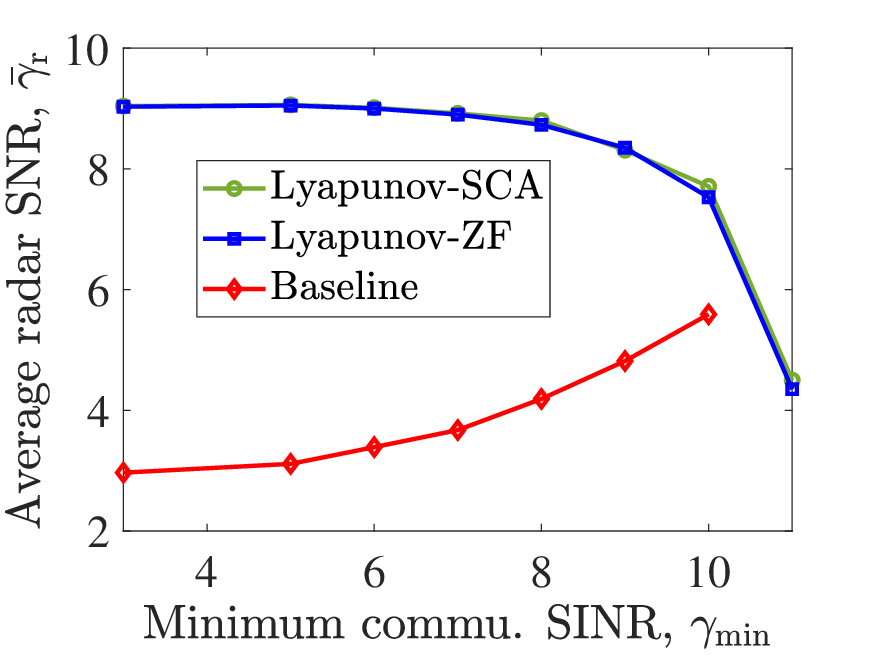}
\label{fig-all_gammamin}
}
\caption{Average radar SNR $\bar{\gamma}_{\textsf{r}}$ vs. the power budget $P_{\max}$ (a), and the minimum average communications SINR limit $\gamma_{\min}$ (b)
}
\label{}
\end{figure}

Fig.~\ref{fig-all_pmax} shows the average radar SNR as a function of the BS power budget $P_{\max}$ (dB) for a fixed minimum average communications SINR limit of $\gamma_{\min} = 10$ dB. The baseline results are plotted from $P_{\max} = 5$ dB, where the algorithm first provides a feasible solution to problem~\eqref{op_1}. The figure highlights the significant performance gains achieved by the proposed dynamic precoding algorithms. Notably, Lya-ZF performs almost identically to Lya-SCA, and as expected, increasing the power budget enhances sensing SNR. However, this improvement is not observed in the baseline algorithm. The reason is that the baseline policy opportunistically allocates more power to the communications subsystem to satisfy the communications SINR constraint, leading to inefficient power usage and, consequently, lower average radar SNR.  

Fig.~\ref{fig-all_gammamin} shows the average radar SNR as a function of the minimum average SINR requirement, given a fixed BS power budget. As the SINR requirement for communications increases, the average radar SNR decreases. This is expected, as the proposed dynamic algorithms allocate more power to communications to meet the required quality of service. However, the decline in radar SNR becomes significant only after a substantial increase in the communications SINR limit, specifically for ${\gamma_{\min} \geq 8}$~dB.  
This suggests that it may be possible to achieve satisfactory radar performance without significantly compromising communications quality of service. Furthermore, Fig.~\ref{fig-all_gammamin} shows that the baseline algorithm becomes infeasible as $\gamma_{\min}$ increases. In such cases, it allocates more power to radar while failing to meet the communications requirement set by $\gamma_{\min}$, leading to an increase in average radar SNR despite the higher communications SINR constraint.

%% file: Chapters/Conclusion.tex
\section{Conclusions }\label{sec_conl}
We developed a dynamic precoding algorithm for a discrete-time JCAS system using Lyapunov optimization. Under the ZF approach, we observed that the optimal precoding at each time slot is to use only one of the communications or the radar precoder. This suggests that the backscattered signal from the communications-intended waves can also be effectively utilized for sensing purposes.
Simulation results demonstrated the effectiveness of our dynamic algorithms. Future research could explore mobility scenarios in communications and time-varying target responses, e.g., moving targets.